\documentclass[journal,12pt]{IEEEtran}

\usepackage{lettrine}
\usepackage[utf8]{inputenc}
\usepackage{cite}
\usepackage{amsmath,amssymb,amsfonts}
\usepackage{algorithmic}
\usepackage{textcomp}
\usepackage{xcolor}
\usepackage{pdfpages}
\usepackage{enumitem}
\usepackage{url} 
\usepackage{multicol,lipsum}
\usepackage[export]{adjustbox}
\usepackage{amsmath,amsfonts,amssymb}
\usepackage{multirow}
\usepackage{booktabs}
\usepackage{cite}
\usepackage{caption}
\usepackage{subcaption}
\usepackage{lipsum}
\usepackage{colortbl}
\usepackage{diagbox}

\ifCLASSINFOpdf
\else
\fi
\usepackage{graphicx}

\begin{document}
\title{Enhancement of Rural Connectivity by Recycling TV Towers with Massive MIMO Techniques}

\author{Ammar~El~Falou,~\IEEEmembership{Member,~IEEE,}
        and~Mohamed-Slim~Alouini,~\IEEEmembership{Fellow,~IEEE}% <-this % stops a space 
        \vspace{-0.78cm}
\thanks{The authors are with King Abdullah University of Science and Technology (KAUST)}}

\begin{titlepage}
\vspace{3cm}
\Huge\noindent\textbf{IEEE Copyright Notice}\\
\vspace{3cm}

\large\noindent\copyright 2022 IEEE. Personal use of this material is permitted. Permission from IEEE must be obtained for all other uses, in any current or future media, including reprinting/republishing this material for advertising or promotional purposes, creating new collective works, for resale or redistribution to servers or lists, or reuse of any copyrighted component of this work in other works.
\end{titlepage}

\maketitle


\begin{abstract}
Nowadays, the digital divide is one of the major issues facing the global community. Around $3$ billion people worldwide are still not-connected or under-connected. In this article, we investigate the use of TV towers with multi user (MU) massive multiple input multiple output (mMIMO) techniques to offer connectivity in rural areas. Specifically, the coverage range is assessed for a MU mMIMO base station (BS) mounted on a high tower as a TV tower, and compared with a legacy mMIMO BS. The obtained results show that one high tower BS can cover an area at least $25$ times larger than the area covered by a legacy BS. This is of high interest as recycling TV towers can enhance the rural connectivity with low expenditures. We apply the proposed solution to a realistic case study in an Ethiopian rural area, based on population densities and locations of current BS and TV towers. Our study shows that a high number of people can be covered by existing TV towers. Non-technological challenges and additional possible solutions to enhance rural connectivity are also discussed. 
\end{abstract}

%\begin{IEEEkeywords}
%TV Towers, Massive MIMO, Rural Connectivity, Digital Divide.
%\end{IEEEkeywords}

% \fi
\IEEEpeerreviewmaketitle

\vspace{-0.6cm}
\section{Introduction}
%\IEEEPARstart{T}{his} demo file is intended to serve as a ``starter file''
%for IEEE journal papers produced under \LaTeX\ using
%IEEEtran.cls version 1.8b and later.
%\vspace{-0.1cm}

\subsubsection*{Digital Divide}
People in developed countries are benefiting from the experience offered by 4G and 5G communication systems in their work, education, health system, social life, and entertainment. On the other side, around $3$ billion people in undeveloped countries and rural areas in some developed countries are still not-connected or under-connected,  a problem globally known as the “digital divide”~\cite{yaacoub2020key,feltrin2021potential}. %,ahmmed2022digital}.
 These regions generally suffer from bad infrastructures with regular power outages and lack of roads and transportation, or have a low density of population. Such reasons have made the cost of deploying cellular sites in rural regions not economically viable. Operators need to account for powering and backhauling costs in their business models and compare them with possible revenues to take deployment decisions. 

%\vspace{-0.3cm}
\subsubsection*{Wireless Connectivity and Super Cells}
Nowadays, providing ubiquitous connectivity is becoming a necessity and is gaining more interest~\cite{yaacoub2020key,osoro2021techno,feltrin2021potential}. %ahmmed2022digital.
 Maintaining a seamless connectivity is mandatory for efficient online education, remote working, e-governance, and online business. To this end, recent efforts in 6G communications are focused on decreasing the digital divide gap to meet future sustainable hyper-connectivity demands~\cite{yaacoub2020key}.

In many non-connected or under-connected areas, wireless connectivity can be seen as the only option for access and backhaul communications~\cite{yaacoub2020key,bondalapati2020supercell}. Indeed, wired communications require high expenses that cannot be justified by the limited economical revenues in rural regions. Several technologies and approaches are proposed for wireless connectivity in rural areas as: high-altitude platform systems (HAPS), satellite communications, unmanned aerial vehicle base station (UAV-BS) and super cells (SCs) (see~\cite{yaacoub2020key,osoro2021techno,bondalapati2020supercell, feltrin2021potential} %,ahmmed2022digital
and references therein). Among them, the technology of SCs, i.e., cells with a high tower and large coverage area can be considered as the most cost-effective ~\cite{yaacoub2020key,bondalapati2020supercell,taheri2021potential,feltrin2021potential}. The advantage of this solution is further amplified when using available TV towers to provide coverage to SCs. 

%The main reasons are: 
%\begin{enumerate}
%    \item Operational expenditure (OPEX) is low especially when towers are already available.
%    \item  high towers provide a propagation advantage as a line of sight (LoS) connection is highly probable.
%    \item Large cells are more suitable for low population areas, as the cell coverage is the main limitation in this case, rather than the capacity as in urban areas. 
%\end{enumerate}  

%\vspace{-0.4cm}
\subsubsection*{Current Situation of TV Transmitters}

Television or TV service appeared at the beginning of the $20$th century, and gained popularity after $1950$s. Terrestrial TV service is typically transmitted via a high tower in order to cover a large area of tens of kilometers. TV towers are connected to the electrical grid and backhauled to broadcast served TV channels. The use of TV towers for extending rural connectivity is mainly motivated by:
\begin{enumerate}[label=(\roman*),noitemsep,nolistsep]
\item The availability of electricity and backhauling in current TV towers thus operational expenditures (OPEX) are lower than other solutions.
\item The high height which helps in providing large coverage areas. Indeed, high towers provide a propagation advantage as a line of sight (LoS) connection is highly probable.
\item Large cells are more suitable for rural areas, as the cell coverage is the main limitation in this case, rather than the capacity as in urban areas.
\item The prior re-allocation of parts of the TV spectrum, mainly the sub-700 MHz band, for communication purposes~\cite{ITU,3GPPNR}. 
\end{enumerate}

For all these reasons, taking full advantage of TV towers for connectivity can be seen as an evident choice to mitigate the digital divide. Note that the TV service remains intact, but its spectrum is efficiently used to permit data communication. 

%\vspace{-0.4cm} 
\subsubsection*{Massive MIMO and 6G Systems} The use of a massive number of antennas for communication known as massive multiple-input multiple-output (mMIMO) largely increase the beamforming gain and enhance the spatial selectivity of multi-user (MU) systems. mMIMO techniques are key technologies to increase the system capacity and coverage, thus improving the user experience~\cite{MassiveMIMOBook}. mMIMO techniques are standard technology in 5G systems, and crucial in beyond 5G systems as 6G.

%~\cite{}.
%Indeed, MIMO precoding techniques aim to: 
%\begin{enumerate}[label=(\roman*),nolistsep,noitemsep]
%    \item Focus the power transmitted by the serving BS on the targeted devices through beamforming, thus increasing the coverage range. 
%    \item Decrease multi-user interference thanks to spatial selectivity and increase the rate of users. 
%\end{enumerate}
%\vspace{-0.4cm}
\subsubsection*{Proposed Work and Contributions}
In this article, we investigate the use of MU mMIMO linear precoding at the top of high towers, as TV towers, to improve the connectivity of rural areas. As a benchmark, we consider a MU mMIMO legacy BS. The user is considered as \textit{connected} in the downlink when its rate exceeds $10$~Mbps, which guarantees a good quality of service to several applications including online education, remote working, and online business~\cite{yaacoub2020key}. By evaluating the \textit{coverage range}, we show that recycled high towers implementing MU mMIMO technology can cover an area at least $25$ times larger than the area covered by a legacy BS. We apply this solution to a \textit{realistic case} study in a low-income country suffering from a high telecommunication service imbalance as in Ethiopia~\cite{zhang2021telecommunication}. We take into consideration the population density maps and the location of present cellular sites and TV towers. We show that a large number of persons can be covered by recycling TV towers for connectivity. The required effective isotropic radiated power (EIRP) per user in the uplink (UL) is also discussed for the Ethiopian case study. Finally, we highlight non-technological challenges and discuss new technologies that can further enhance the connectivity in rural areas such as: low-earth orbiting (LEO) satellites, non-orthogonal multiple access (NOMA) and hybrid mMIMO.

In the literature, similar works are listed as:
\begin{itemize}[nolistsep,noitemsep]
%\subsubsection{Meta Engineering Super Cell} 
\item \textit{Meta Engineering Super Cell}\cite{bondalapati2020supercell}: SCs are proposed in \cite{bondalapati2020supercell} by Meta Connectivity as a large-area coverage cell that leverages towers up to $250$~m height, equipped with high-gain, narrow-sectored antennas in order to increase the coverage range and the capacity. Drive tests were used to assess the advantage of this solution. A 36-sector SC with Luneberg Lens is deployed and it is shown that one SC can replace 15 to 25 traditional macrocells. Single outdoor user is considered in the drive tests.   

\item \textit{Curvalux}\cite{Curvalux}:  Curvalux  proposed a phased array multi-beam antenna to be installed on legacy cellular towers. This array forms 16 high-gain beams and cover a sector having an angle equal to 60$^{\circ}$. Many proof-of-concept (POC) trials have been done in several countries as Philippines, Mongolia, Indonesia, Kuwait. These POC validate the feasibility of the proposed multi-beam antenna but technical details and measurement are not published. 
\item \textit{Ericsson} \cite{feltrin2021potential}: Ericsson researchers studied high towers with large antenna arrays to deliver long range connections. They discussed the performance of satellite and/or terrestrial large cell systems, based on the traffic density and required infrastructure, to offer rural connectivity. Simulation results have shown that the two solutions deliver connectivity in complementary traffic and partly different scenarios.
 
\item \textit{Luleå University} \cite{taheri2021potential}: The potential of mMIMO on TV towers for cellular coverage extension has been investigated. The Ericsson 9999 rural path-loss model and independent Rayleigh fading are considered. Different carrier frequencies have been considered mainly 700 MHz, 1800 MHz and 3500 MHz and a bandwidth of 20 MHz is used with all frequencies. A single-cell BS, operating in a time division duplex (TDD) mode, with a big size mMIMO uniform cylindrical array (UCyA) of $2$~m radius and up to $10$~m height, is considered. Under these assumptions, authors in~\cite{taheri2021potential} have shown that an area of radius $70$~km can be covered around a TV tower using the mMIMO technology.
\end{itemize}

The main differences between our work and previous ones are: \begin{enumerate}[noitemsep,nolistsep]
    \item In~\cite{feltrin2021potential,bondalapati2020supercell, Curvalux} fixed beam high-gain antennas are used regardless of the location of users. This solution might not be energy-efficient. Moreover, the performance of a single user is assessed without considering intra-cell interference. In our work and in~\cite{taheri2021potential}, MU mMIMO beamforming is employed, adapting based on the channel state information (CSI) of served users. In this case, CSI estimation is necessary.
    \item In~\cite{taheri2021potential}, a zero-forcing (ZF) linear precoder is considered which might not be efficient when users are close to each other. %\cite{ratnayake2012effects}. 
    We consider a regularized ZF (RZF) precoder which is more suitable than ZF in practice, with limited added complexity and cost~\cite{MassiveMIMOBook}.   
    \item In our work, we consider the 3GPP channel models \cite{3GPP} based on realistic measurements rather than the Ericsson 9999 model. These models are generated using the open source geometry-based quasi deterministic radio channel generator (QUADRIGA) stochastic channel simulator v.2.6~\cite{quad}. Simulations are done with different types of 3GPP 38.901 models mainly line of sight (LoS) for high tower BS and non LOS (NLoS) for legacy BS. On the other hand, authors in~\cite{taheri2021potential} use Ericsson 9999 model with independent Rayleigh fading channel. This assumption is not realistic especially in the case of high towers where LoS is paramount.    
    \item Authors in \cite{taheri2021potential} consider a mMIMO BS with a large UCyA that can accommodate up to $293 \times 464 = 135,952$ antennas. Digital precoding, e.g., ZF requires as much radio-frequency (RF) chains as antennas. The implementation of this number of RF chains is not possible. Moreover, the implementation of a UCyA with a radius of $2$~m and a height of $10$~m might be challenging to resist against wind in practice.
    \item Different from \cite{taheri2021potential}, we complement our work by taking into consideration a real case study in Ethiopian rural area, with population density and location of legacy BSs and TV towers. We differentiate between three cases: low, medium, and high number of served users per cell. %Note that in \cite{taheri2021potential}, the population density per km$^2$ $p$ is considered equal to $0.15$ while it is
\end{enumerate}
Our study is important to motivate network operators to recycle TV towers for rural connectivity.
%The novel 
%Difference:
%comparative 
%more users

%In this paper, we consider long-range rural wireless communications. We focus on the access link and backhaul-link.  

%Two parts:
%\begin{enumerate}
%    \item Point-to-point: Direct access to users 4G, 5G; frequencies 
%    \item Wireless Back-haul (TV spectrum), Long Haul Wireless Back-Haul
%    \item Validate the channel model with Facebook measurement (Beyond 10 Km)
%    \item Modeling
%\end{enumerate}

%different rural geographical environments and different renewable energy solution (solar, wind, etc.)

%hybrid MIMO

%Cell-free MIMO ?

%energy efficiency

%Channel estimation ?

%Coverage with a target throughput form standard ? 10 Mbps \cite{yaacoub2020key}

%wind resistance?

%Hybrid massive MIMO and LEO networks?

%RIS-assisted MIMO (passive RIS) ?

%NOMA ?

%Comparison with HAPS, UAV, (table in FB paper)?

%draw coverage based on percentage (with + without LEOS)

 

%\subsection{Subsection Heading Here}
%Subsection text here.

%\IEEEpubidadjcol

%\subsubsection{Subsubsection Heading Here}
%Subsubsection text here.

\vspace{-0.45cm}
\section{System Model}
\vspace{-0.1cm}
\subsection{DL MU Massive MIMO Base Stations}
%\subsection{Channel estimation}
\begin{table*}
\normalsize \centering 
\caption{System Parameters} 
\vspace{-0.1cm}
%of the running example. Each cell covers a square area of $1$ km $\times 1$ km and is deployed on a grid of $4 \times 4$ cells. A wrap-around topology is used. The devices are uniformly and independently distributed in each cell, at distances larger than 35m from the BS.}
\setlength\tabcolsep{1.5pt}
\begin{tabular}{|c|c|c|} 
\hline
Parameter & Legacy BS & High Tower BS (TV Tower) \\
\hline
\multicolumn{3}{|c|}{} \\[-13.5pt]
\hline
Tx Height [m]    & $25$ & $150$ \\  
\hline
Rx Height [m]    & $8$ & $8$ \\  
\hline
Channel Model     & 3GPP 38.901 RMa NLOS & 3GPP 38.901 RMa LOS \\       
\hline 
Nb. of Active Users $K$  & $20,50,100$ & $ 20,50,100$ \\
\hline
Target DL Rate [Mbps] & $10$ & $10$  \\
\hline
Number of Antennas     & $256$ single/$512$ dual polarized & $256$ single/$512$ dual polarized \\
\hline
 $M_h \times M_v$  & $32 \times 8$ (single)/$32 \times 8\times2$ (dual) & $32 \times 8$ (single)/$32 \times 8\times2$ (dual)\\
\hline
Bandwidth $B$ [MHz] &  $10, 20, 100$ & $10, 20, 100$ \\ 
\hline
Transmit Power [dBm] & $46$  & $50$\\
\hline
Cyclic Prefix Overhead  & $5\%$ & $5\%$ \\
\hline
\multicolumn{3}{|c|}{} \\[-13.5pt]
\hline
Carrier Frequency $f_c$ [MHz] & \begin{tabular}{lcr} $700$ & $1800$ & $3500$  \end{tabular} & \begin{tabular}{lcr}$700$&$1800$&$3500$ \end{tabular} \\
\hline
Height $h_a$ [m] & \begin{tabular}{lcr} $1.50$ & $0.59$ & $0.30$ \end{tabular} & \begin{tabular}{lcr} $1.50$ & $0.59$ & $0.30$  \end{tabular}\\
\hline
Radius $r_a$ [m] & \begin{tabular}{lcr} $1.09$ & $0.42$ & $0.21$ \end{tabular}  & \begin{tabular}{lcr} $1.09$ & $0.42$ & $0.21$ \end{tabular}\\
\hline 
\end{tabular}
\label{tab:SystemParmeters}
\vspace{-0.6cm}
\end{table*}

\begin{figure}
\vspace{-0.54cm}
    \centering
\includegraphics[width=0.92\columnwidth]{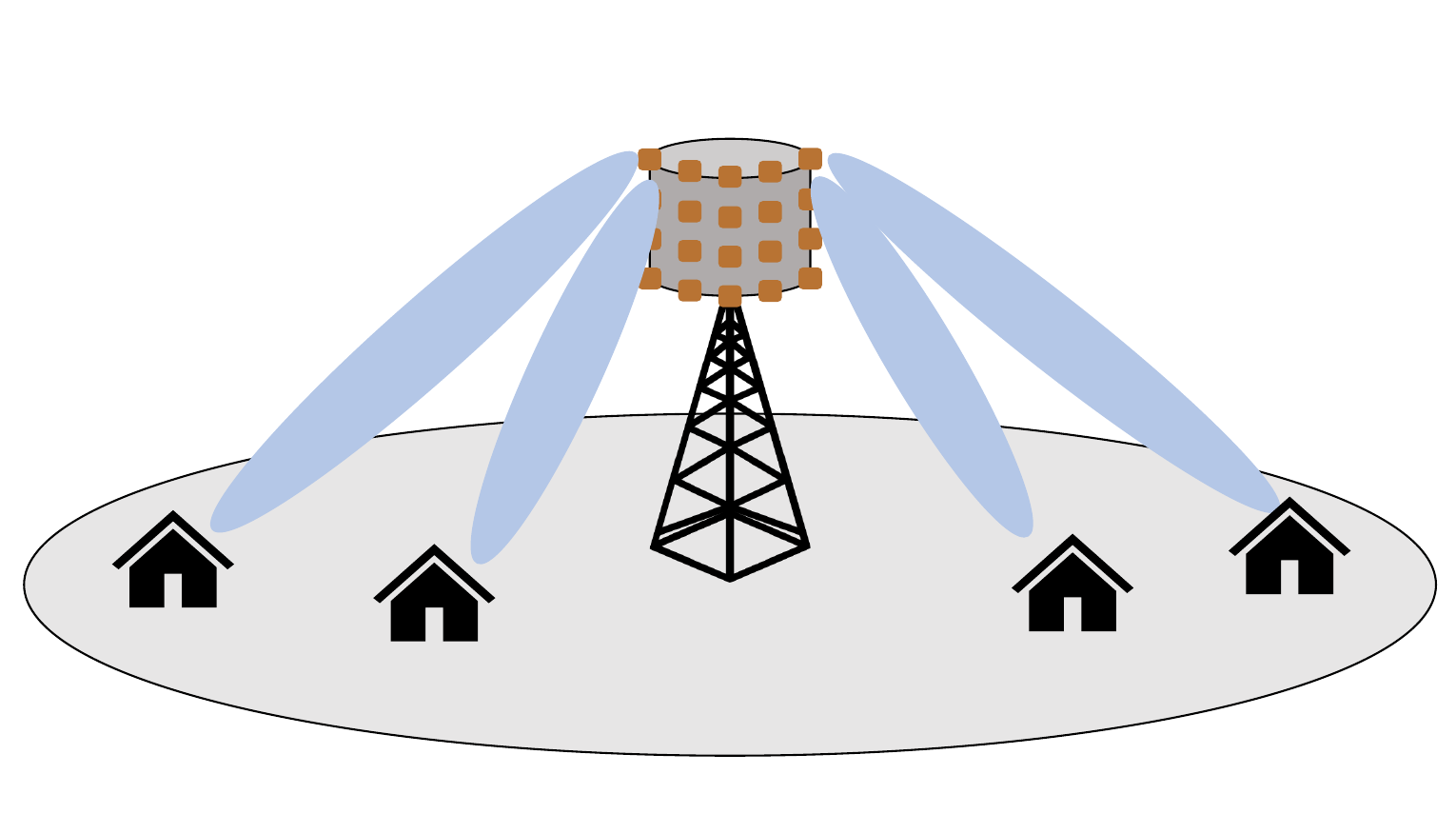}
    \caption{System model: A mMIMO base station.}
    \label{fig:SystemModel}
  \vspace{-0.75cm}
\end{figure}

We consider a downlink (DL) MU mMIMO single-cell BS with $256$ single-polarized antennas (or $512$ dual-polarized antennas) forming a UCyA as depicted in Fig.~\ref{fig:SystemModel}. The number of horizontal antennas $M_h = 32$ and vertical antennas $M_v = 8$. The system parameters are listed in Table~\ref{tab:SystemParmeters}. The increase of the number of antennas in a direction will increase its beamforming gain. In order to have a system with a reasonable size and complexity, we restricted the number of antennas. We note that in practical scenarios, antennas in the azimuth can help in the separation of users in the horizontal direction and antennas in the elevation can help in the separation of users in the vertical direction. For rural areas, it is preferable to have antennas in the azimuth more than in the elevation as users are generally distributed in the horizontal direction. 

According to \cite{yaacoub2020key}, we consider a target rate per user of 10 Mbps. To ensure fairness between users, a max-min power control is done maintaining similar signal to interference plus noise ratio (SINR) values for all users~\cite{MassiveMIMOBook}. We assume that the receiver is placed at a height of~$8$~m. This can represent the case when the receiver antenna is placed at the roof of the house unit in rural areas. An omnidirectional antenna is considered. An antenna with certain gain at the receiver will help in increasing: either the BS coverage range or the data rate of its respective user.

The number of active users per cell $K \in \{20, 50, 100\}$ reflects a low, medium, and high number of active users distributed uniformly around the BS. The cyclic prefix overhead for orthogonal frequency division multiplexing (OFDM) is considered at $5\%$. Three carrier frequencies in the sub-$6$~GHz are considered: $700$, $1800$, and $3500$~MHz. The corresponding bandwidths are: $10$~MHz, $20$~MHz, and $100$~MHz as in the 5G NR frequency bands~\cite{3GPPNR}. Note that systems using the former frequencies, $700$ and $1800$~MHz, operate in frequency division duplex (FDD) mode where the full bandwidth is used in DL, while those using the later frequency, $3500$~MHz, operate in a TDD mode where $3/4$ proportion of the bandwidth is for DL. For mMIMO systems, TDD mode is preferred as the channel estimation scales with the number of users and not with the number of antennas. However, for sub-$6$~GHz frequencies, the number of antennas is relatively low and both FDD and TDD are used~\cite{3GPPNR}. The RZF linear precoder is used as an efficient and low-complex precoder for MU MIMO systems. Other precoders such as: ZF, maximum ratio (MR) or minimum mean-squared error (MMSE) are not recommended. Indeed, ZF is not efficient when users are close to each other. MR is low-complex but has a limited data rate while MMSE is optimal but very complex for practical implementation. A perfect CSI is assumed. Fortunately, RZF performs well even with channel estimation using low complex estimators as least-squares (LS)~\cite{MassiveMIMOBook}.\\
Two type of single-cell BSs are considered:
\begin{itemize}[noitemsep,nolistsep]
    \item High tower BS (TV tower): A BS with a transmitter height of $150$~m and a transmit power of $100$~W. The value of the tower height is inspired from digital terrestrial television broadcasting (DTTB) where TV tower effective height is equal to $150$~m in rural scenarios~\cite{ITU}.  
    \item Legacy BS: A BS with a transmitter height of $25$~m and a transmit power of $40$~W. These parameters are considered as a benchmark. 
\end{itemize}
\vspace{-0.61cm}
\subsection{Antenna Array and Channel Models}
\begin{table*}
\normalsize \centering 
\caption{Coverage Distance for a Legacy BS and a High Tower BS (TV Tower)} 
\vspace{-0.1cm}
\setlength\tabcolsep{1.5pt}
\begin{tabular}{|c|c|c|c|c|c|c|} 
\hline
\rule{0pt}{10pt}    
 Type & $K$ & $f_c$ [MHz] & Duplex Mode & $B$ [MHz] & $d_\text{cov}$ single [Km] & $d_\text{cov}$ dual [Km]  \\
\hline
\multicolumn{7}{|c|}{} \\[-13.5pt]
\hline
\multirow{9}{*}{Legacy BS} & 20 &  &  & & 4.1 & 4.9 \\  
\cline{2-2}\cline{6-7}
 & 50 & 700 & FDD & 10 & 2.7 & 3.4 \\  
\cline{2-2}\cline{6-7}
 & 100 &  &  &  & 1.7 &  2.1\\  
\cline{2-7}
 & 20 &  &  &   & 2.9 &3.5\\  
\cline{2-2}\cline{6-7}
 & 50 & 1800 & FDD & 20 & 1.9 &2.5\\  
\cline{2-2}\cline{6-7}
 & 100 &  &  &  & 1.2 & 1.6\\  
\cline{2-7}
 & 20 &  &  &  & 2.2 &  2.7 \\  
\cline{2-2}\cline{6-7}
 & 50 & 3500 & TDD (DL: 3/4) & 100 & 1.5 &  1.9 \\  
\cline{2-2}\cline{6-7}
& 100 &  & &  & 1 &   1.4  \\  
\hline
\multicolumn{7}{|c|}{} \\[-13.5pt]
\hline
\multirow{9}{*}{\begin{tabular}{c} High Tower BS \\ (TV Tower)  \end{tabular}} & 20 &  & &  & 28 &37 \\ 
\cline{2-2}\cline{6-7}
 & 50 & 700 & FDD & 10 & 15 & 21 \\  
\cline{2-2}\cline{6-7}
 & 100 &  &  &  & 9.5 & 12.5 \\  
\cline{2-7}
 & 20 & & & & 14.5 & 16.5\\  
\cline{2-2}\cline{6-7}
 & 50 & 1800 & FDD & 20 & 10.5 & 13 \\  
\cline{2-2}\cline{6-7}
 & 100 & &  &  & 7.5 & 9.5 \\  
\cline{2-7}
 & 20 & & &  & 13.5 & 15\\  
\cline{2-2}\cline{6-7}
 & 50 & 3500 & TDD (DL: 3/4) &100 & 10 &  13\\  
\cline{2-2}\cline{6-7}
 & 100 &  & & & 8 & 10 \\  
%\hline
%Channel gain at 1 km   & $\Upsilon = -148.1 $ dB \\
%\hline
%Pathloss exponent     & $ \alpha = 3.76$  \\
%\hline
%Shadow fading (standard deviation)   & $\sigma_\text{sf} = 10$ \\ 
%\hline 
%Carrier frequency   & $f_c = 2$ GHz\\ 
%\hline 
%Receiver noise power   & $-94$ dBm \\ 
%\hline 
%DL transmit power per device   & $100$ mW \\ 
\hline 
%Samples per coherence block   & $\tau_c=2000$  \\ 
%\hline 
%Pilot reuse factor   & $f=1$ \\ 
%\hline 
\end{tabular}
\label{tab:SystemDistance}
\vspace{-0.52cm}
\end{table*}
We use in this work two 3GPP channel models for rural areas \cite{3GPP}, mainly, the 3GPP rural macro NLoS (3GPP  $38.901$ RMa NLoS) for the legacy BS and the 3GPP rural macro LoS (3GPP  $38.901$ RMa LoS) for the high tower BS. \textcolor{black}{The main difference in performance is due to the use of NLoS/LoS models for each type of BS.} Note that these models are based on real measurements. In our simulations, we generate the channel coefficients using the QUADRIGA channel simulator v.2.6~\cite{quad}. The BS is placed in the center of the cell and the UCyA is placed at the top of the tower. The separation between antennas is equal to $\lambda/2$ where $\lambda$ is the wavelength. The use of 3GPP rural channel models and QUADRIGA will make the presented simulations more realistic as they account for several geometric aspects of the system as users locations, correlation between antennas, and heights. Note that 3GPP models are proposed for distances $\leq 10$~km. We did not notice any anomaly for higher distances.
\vspace{-0.55cm}
\subsection{Coverage Range Results}
We conducted extensive Monte-Carlo simulations in order to compute the coverage range of the two types of BSs. The coverage range is defined as the range for which more than $95\%$ of users have a rate higher than $10$~Mpbs. The results are presented in Table~\ref{tab:SystemDistance} where $d_\text{cov}$ stands for the coverage distance, or the maximum value of the coverage range. It is clear in Table~\ref{tab:SystemDistance} that the coverage distance is multiplied by more than $5$ when using a high tower BS instead of a legacy BS with the same mMIMO configuration. This reflects that one high tower BS can cover a $25$~times bigger area than a legacy BS. This result is consistent with the measurements done in~\cite{bondalapati2020supercell} when studying super cells. Note that for a low number of active users, e.g., $K=20$, and a low carrier frequency $f_c = 700$~MHz, this ratio can go up to $57$ times with dual-polarized antennas. 
%Integrated access-backhaul ?
\vspace{-0.5cm}
\section{Mapping to a real case}
\label{sec:covdist}

\begin{figure*}
    \vspace{-0.2cm}
    \centering
    \begin{subfigure}[b]{0.95\columnwidth}
        \centering
        \includegraphics[width=0.9\textwidth]{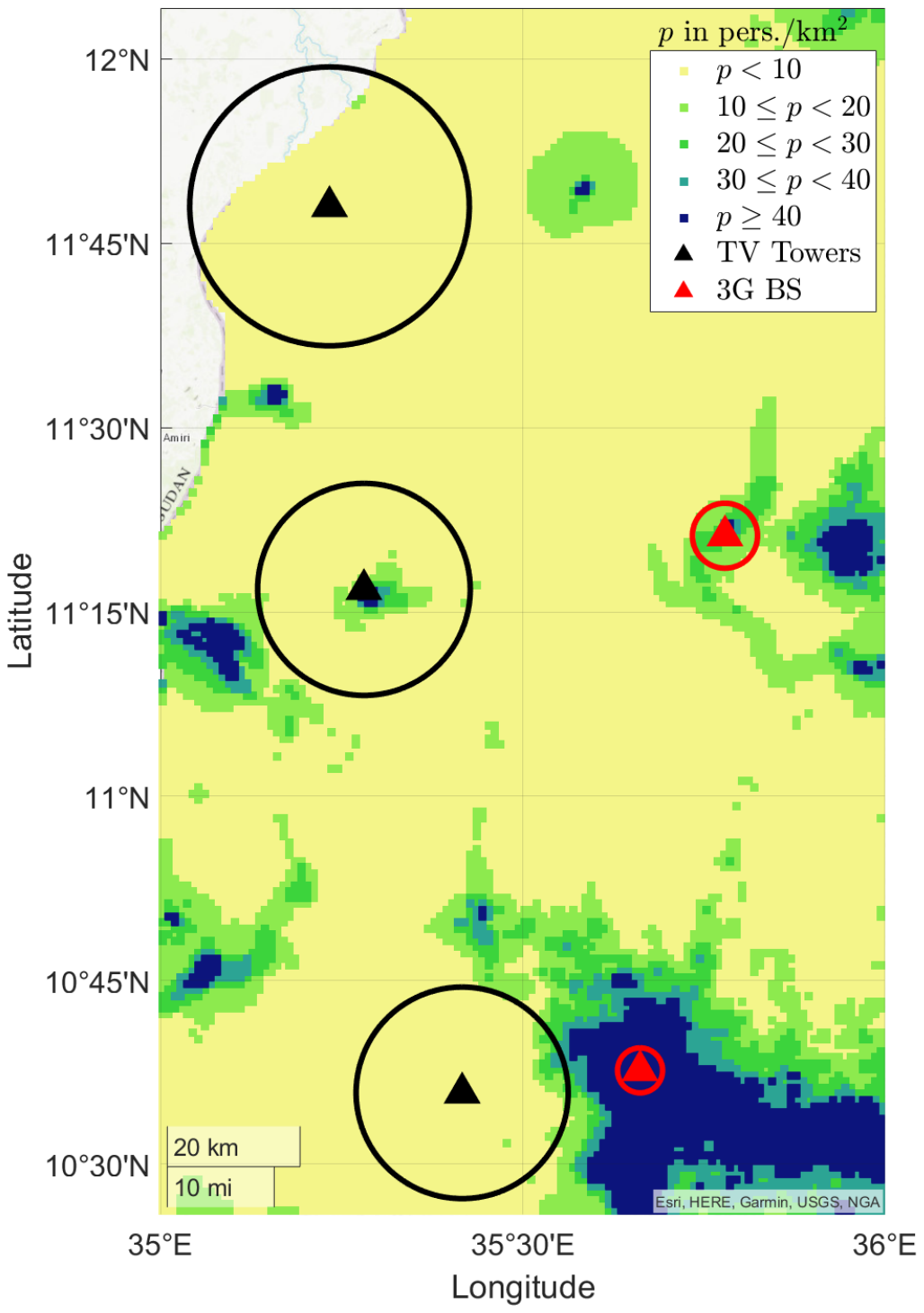}
        %\caption{Population density per km$^2$ $p$}
         \vspace{-0.6cm}
        \caption{Current TV towers}
        \label{fig:EthRuralCov}
    \end{subfigure}
    \hfill
    \begin{subfigure}[b]{0.95\columnwidth}
        \centering
        \includegraphics[width=0.9\textwidth]{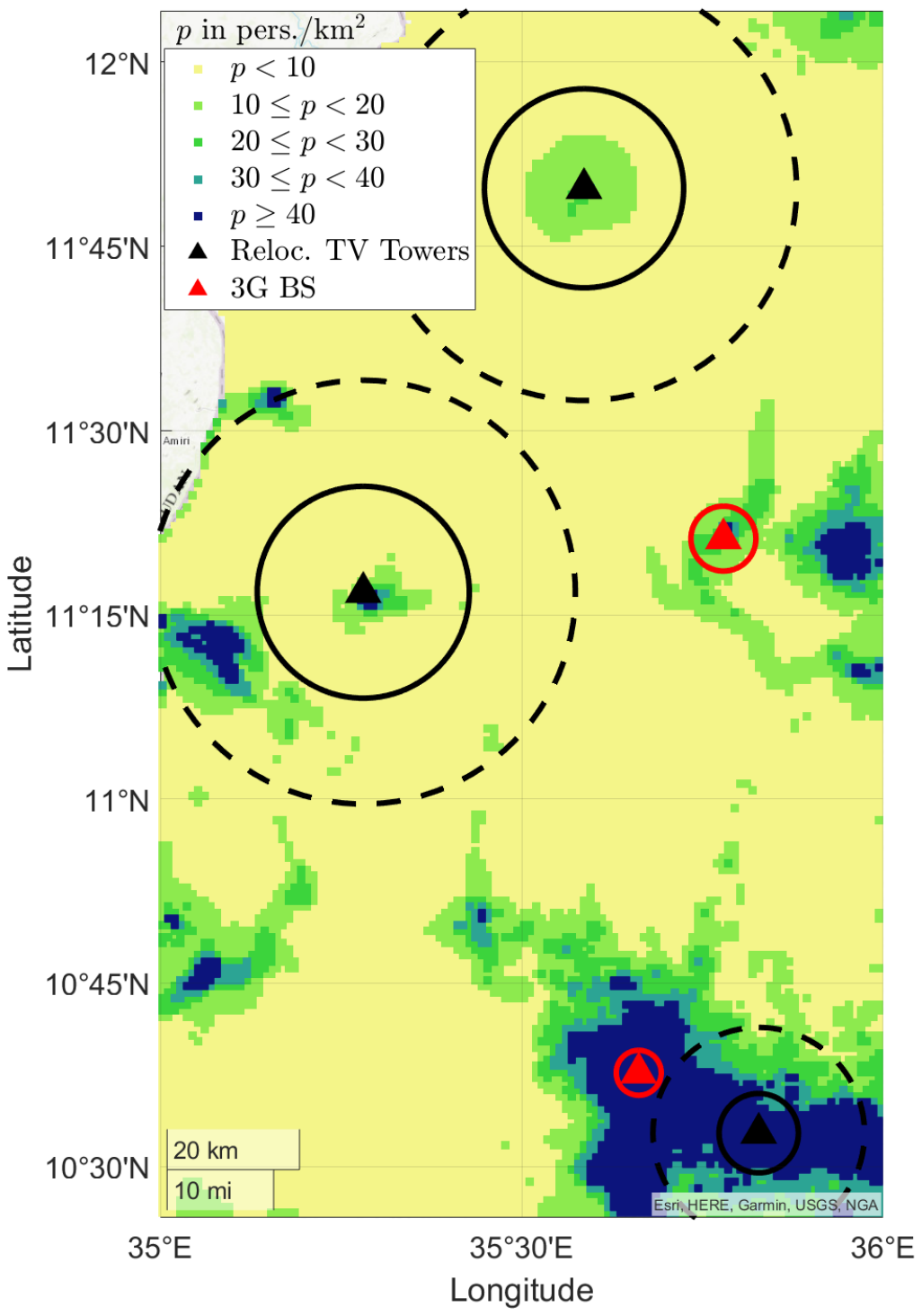}
         \vspace{-0.6cm}
        \caption{Relocated towers}
        \label{fig:EthRuralCovReloc}
    \end{subfigure}
     \vspace{-0.15cm}
    \caption{Rural area in Ethiopia: population density per km$^2$ $p$ and coverage areas.}
    \label{fig:RuralEthiopia}
  \vspace{-0.5cm}
\end{figure*}

% \begin{figure}
%     \centering
%     \includegraphics[width=0.73\columnwidth]{EthRuralwithCov.eps}
%     \caption{Current TV towers: $p$ and coverage areas.}
%     \label{fig:EthRuralCov}
%   \vspace{-0.8cm}
% \end{figure}
In this section, we map the obtained results to the case of a rural area in Ethiopia. We consider the area depicted in Fig.~\ref{fig:EthRuralCov}. This rural area is at the borders of Ethiopia, specifically between latitude: $10^{\circ}30'$N- $12^{\circ}$N, and longitude $35^{\circ}$E-$36^{\circ}$E. Fig.~\ref{fig:EthRuralCov} shows the population density per km$^2$, denoted by $p$, and the location of legacy cellular BSs and TV towers inside this area. The population density is obtained from Facebook/Meta density population data~\cite{Facebook}. The total number of people in this area is about $200,000$ people. The locations and types of cells (3G/4G) are obtained from~\cite{celldata} and the locations of TV towers are obtained from \cite{TVdata}. From Fig.~\ref{fig:EthRuralCov}, it is clear that:
%\begin{enumerate}[noitemsep,nolistsep] 
1) A large part of the considered area is not covered by any 3G BS. 2) Legacy BSs are located in some places with high population density. 3) 4G BSs are not available in the considered area. 4) Some TV towers are available in the considered area.
%\end{enumerate}

% \begin{figure}
%     \centering
%     \includegraphics[width=0.73\columnwidth]{EthRuralwithCov_Relocatedv2.eps}
%     \caption{Relocated towers: $p$ and coverage areas.}
%     \label{fig:EthRuralCovReloc}
%   \vspace{-0.8cm}
% \end{figure}
In order to assess the benefit of recycling TV towers for connectivity, we also plot in Fig.~\ref{fig:EthRuralCov} the coverage circles around legacy BSs and TV tower BSs when both use MU mMIMO. The favored frequency for propagation, i.e., $f_c = 700$~MHz is considered. The percentage of active users from the population is between $1\%$ and $2\%$, identified as a loaded scenario for rural areas~\cite{feltrin2021potential,osoro2021techno}. In practice, this value depends on the adoption rate of Internet service, the served traffic per user, and the percentage of active users. \textcolor{black}{Note that temporal traffic aspects are note considered in our work.} From Fig.~\ref{fig:EthRuralCov}, we can conclude that a substantial increase in the covered area is seen by recycling TV towers. Indeed, the actual cellular network can cover only $9,000$~persons out of the total population of $200,000$ even when using MU mMIMO. This represents $4.5\%$ of the total population. By recycling TV towers for communication, $19,500$ more persons can be covered in the considered scenario, and the percentage of covered persons reaches $14.25\%$. 

Furthermore, it is clear from Fig.~\ref{fig:EthRuralCov} that two TV towers out of the three available ones are covering areas with a low population density, i.e., $p<10$ persons per km$^2$. If a relocation of the TV towers is possible, the number of covered persons by high towers can be increased to $30,000$ as shown in Fig.~\ref{fig:EthRuralCovReloc} using coverage circles with solid lines. The percentage of covered persons then becomes $19.5\%$. We highlight that the relocation of TV towers will induce CAPEX and the amount is to be compared with the cost of deploying new high towers.

For the early adoption of Internet in rural areas, the number of covered persons can largely increase and the percentage of covered persons can reach $63\%$. The corresponding coverage areas are shown in Fig.~\ref{fig:EthRuralCovReloc} with dashed lines. This early adoption corresponds to the case where $2\%$ of persons subscribe for Internet service (adoption rate) while $1/20$ out of them are active at a certain time, as assumed in~\cite{osoro2021techno} in their study on the engineering-economics aspects of satellite broadband constellations for connectivity purposes. The optimization of the location of high towers and their parameters as: transmit power, number of antennas, and height, can further increase the coverage percentage. In practice, a duality exists between capacity and coverage. The use of large towers in villages and cities with high adoption rates is of little importance. For urban areas, connectivity is better provided with legacy BSs.

\begin{table}
\normalsize \centering 
\color{black}
\caption{UL achievable rate per user in Mbps.}
\vspace{-0.1cm}
\setlength\tabcolsep{1.5pt}
\begin{tabular}{|c|c|c|c|c|c|c|} 
\hline
\rule{0pt}{10pt}    
 Type & $K$ &\backslashbox[18mm]{$d_{cov}$}{EIRP}  & 40 & 33 & 30 & 23  \\
\hline
\multicolumn{7}{|c|}{} \\[-13.5pt]
\hline
\multirow{3}{*}{Legacy BS} & 20 & 4.9 & 27 & 12 & 8 &  2.3 \\
\cline{2-7}
 & 50 & 3.4 & 30  & 14.5 & 10  & 3.2 \\  
\cline{2-7}
 & 100 & 2.1& 39 & 21 & 15 &  5.7 \\  
\hline
\multicolumn{7}{|c|}{} \\[-13.5pt]
\hline
\multirow{3}{*}{\begin{tabular}{c} High Tower BS \\ (TV Tower)  \end{tabular}} & 20 & 37 & 12.5 & 4.5 & 2.5 &  0.85 \\ 
\cline{2-7}
 & 50 & 21 & 16.5 & 7 & 4.5 & 1.55 \\  
\cline{2-7}
 & 100 & 12.5 & 22 & 10 & 6.5 & 2.5 \\ 
%\hline 
%Pilot reuse factor   & $f=1$ \\ 
\hline 
\end{tabular}
\label{tab:UL}
\vspace{-0.7cm}
\end{table}

To provide a more comprehensive study, we evaluated the performance in the UL. In Table~\ref{tab:UL}, we provide the UL achievable rate per user, by more than 95\% of users, for several EIRP values, with $f_c \!=\! 700$ MHz and users uniformly distributed around the BS in the same range as in Table~\ref{tab:SystemDistance}. Practical systems with realistic analog to digital converters require a limitation of the received power ratio $\delta$ between weakest and strongest users  \cite{MassiveMIMOBook}. A power ratio of $\delta \!=\! 20$ dB is considered.  It is clear that UL rates for EIRP $\!=\! 40$ and $\!33\!$ dBm are acceptable and the number of covered persons can be maintained. The rate per user increases with the number of users $K$ as: 1) $d_\text{cov}$ decreases then the path loss decreases, 2) the number of antennas at the BS remains largely higher than $K$ allowing for intra-cell interference cancellation. The UL rate for EIRP $\!=\! 30$ dBm might be acceptable depending on the application. However, the rate for EIRP $\!=\!23$ dBm is low. High EIRP values are obtained by increasing the user transmit power and/or using high gain antennas. In our work, considering high gain antenna is reasonable as it is placed at the rooftop.

\vspace{-0.6cm}
\section{Non-technological Challenges}
In addition to the technological factors, one can mention several non-technological challenges as~\cite{yaacoub2020key,feltrin2021potential}: i) Lack or absence of electricity forcing telecommunication operators to deploy their own diesel generators, ii) bad road infrastructure making the transportation of telecommunication and electrical equipment complicated, iii) Lack or absence of skilled engineers for deployment and maintenance. All these factors increase both CAPEX and OPEX making the deployment and operation of telecommunication systems in rural areas more difficult.  
%\begin{figure}
%    \centering
%    \includegraphics[width=0.96\columnwidth]{MassiveMIMOSystemModelExtended.pdf}
%    \caption{Possible enhancements to the system.}
%    \label{fig:SystemModelExtended}
%   \vspace{-0.4cm}
%\end{figure}
\vspace{-0.6cm}
\section{Possible enhancements}
\label{sec:enhancements}
In this article, we have shown that recycling TV towers for enhancing rural connectivity is a promising solution. However, this solution is not sufficient to cover the entire rural areas. Possible enhancements include:
\subsubsection*{LEO Satellites}
LEO satellites can be used to cover rural areas that are not reachable by the terrestrial towers. Indeed, LEO constellations can provide low-latency high-rate communication links for rural users \cite{feltrin2021potential,osoro2021techno}. In practice, LEO constellations are more suitable for rural areas in rich developed countries than undeveloped poor countries. Their high deployment cost requires expensive subscription fees to make their business model viable. mMIMO can be used with LEO satellites to increase the beamforming gains%~\cite{you2020leo}
, and decrease the OPEX in order to reduce the subscription fees.
%\vspace{-0.4cm}
\subsubsection*{NOMA}
In the presented results, we assumed that active users are uniformly distributed around the cell tower. In reality, persons, and consequently users, tend to live close to each others as shown in~Fig.~\ref{fig:RuralEthiopia}, which increase multi-user interference. Multiple access techniques should be used in order to mitigate this interference. NOMA is %~\cite{khaled2021multi}
preferred over OMA as it can offer higher data rates. NOMA allows the service of two or more users using the same beam. For long distances as in our work, power allocation strategies for NOMA users are not complex and their separation can be easily done using interference cancellation techniques. 
\subsubsection*{Hybrid MIMO} In order to increase the beamforming gain and the coverage range, the number of transmit antennas should be increased at the BS. Digital precoding requires as many RF chains as antennas. These additional RF chains increase the cost and complexity of the system. Hybrid MIMO is a possible way to increase the number of antennas while maintaining the cost and complexity of the system at acceptable levels. Hybrid MIMO and NOMA can be combined together to further enhance the performance of the system.%~\cite{khaled2021joint}.

\vspace{-0.5cm}
\section{Conclusion}
In this article, we have shown that recycling TV towers with mMIMO precoding provides a low-cost enabler of connectivity in rural areas. Indeed, one high tower BS can cover an area at least $25$ times larger than a Legacy BS. Furthermore, we highlighted the benefits of the proposed solution in a realistic use case (Ethiopian rural area) where we showed that a large number of people can be covered by deploying a BS at the top of available TV towers. 
%Furthermore, we have discussed some future directions to enhance the performance of the proposed solution.
An important perspective of this work is to consider the terrain information (buildings, valleys, trees, mountains, hills) in addition to the population density maps and towers locations in the coverage study. This will permit drawing a coverage map similar to our proposed one where network operators can obtain directly the number of possible subscribers in a region and include this number in their business model.

\ifCLASSOPTIONcaptionsoff
  \newpage
\fi



%\bibliographystyle{IEEEtran}

\vspace{-0.2cm}
\bibliographystyle{IEEEtran}
\bibliography{IEEEabrv,Bibliography}

%\begin{thebibliography}{1}


%\end{thebibliography}

\vspace{-1.3cm}
\begin{IEEEbiographynophoto}%[{\includegraphics[width=1in,height=1.25in,clip,keepaspectratio]{AmmarElFalou.png}}]
{Ammar El Falou [M]}  (ammar.falou@kaust.edu.sa) received his Ph.D. degree in communication and information science from IMT Atlantique, France, in 2013. He is currently a Research Engineer at King Abdullah University of Science and Technology (KAUST), Saudi Arabia. His current research interests include massive MIMO systems, rural connectivity and wireless communication systems.
\end{IEEEbiographynophoto}
\vspace{-1.25cm}
\begin{IEEEbiographynophoto}%[{\includegraphics[width=1in,height=1.25in,clip,keepaspectratio]{Alouini.png}}]
{Mohamed-Slim~Alouini [F]}  (slim.alouini@kaust.edu.sa) received his Ph.D. degree in electrical engineering from the California Institute of Technology, Pasadena, in 1998. He is now a Distinguished Professor of Electrical Engineering at KAUST. His current research interests include the modeling, design, and performance analysis of wireless communication systems.
\end{IEEEbiographynophoto}
\end{document}